\documentclass[11pt,a4paper]{article}

\usepackage{amsmath}
\usepackage{amsthm,amssymb}

\usepackage{graphicx} \graphicspath{{figs/}}

\newcommand{\Ai}{\operatorname{Ai}}
\newcommand{\w}{\omega}
\newcommand{\s}{\sigma}
\newcommand{\const}{\mbox{const\ }}
\newcommand{\into}{\rightarrow}
\newcommand{\nn}{\nonumber \\}
\renewcommand{\l}{\lambda}

\numberwithin{equation}{section}

\begin{document}
\title{Exact Solution of Semi-Flexible and Super-Flexible Interacting
  Partially Directed Walks}
\author{A L Owczarek$^1$ and T Prellberg$^2$\\
  \footnotesize
  \begin{minipage}{13cm}
    $^1$ Department of Mathematics and Statistics,\\
    The University of Melbourne, Victoria~3010, Australia.\\
    \texttt{a.owczarek@ms.unimelb.edu.au}\\[1ex] 
$^2$ School of Mathematical Sciences\\
Queen Mary, University of London\\
Mile End Road, London E1 4NS, UK\\
\texttt{t.prellberg@qmul.ac.uk}
\end{minipage}
}

\maketitle  

\begin{abstract}
  We provide the \emph{exact} generating function for semi-flexible
  and super-flexible interacting partially directed walks and also
  analyse the solution in detail. We demonstrate that while fully
  flexible walks have a collapse transition that is second order and
  obeys tricritical scaling, once positive stiffness is introduced the
  collapse transition becomes first order.  This confirms a recent
  conjecture based on numerical results. We note that the addition of
  an horizontal force in either case does not affect the order of the
  transition. In the opposite case where stiffness is discouraged by
  the energy potential introduced, which we denote the super-flexible
  case, the transition also changes, though more subtly, with the
  crossover exponent remaining unmoved from the neutral case but the
  entropic exponents changing.

 \end{abstract}
\section{Introduction}

The collapse transition of an isolated polymer has continued to
attract both theoretical and experimental attention. The canonical
lattice model of single polymer collapse has been the self-avoiding
walk with the addition of attractive potentials between non-bonded
nearest-neighbour sites of the walk. This is known as the Interacting
Self-avoiding Walk (ISAW). This model has yielded many important
theoretical aspects of the physical problem though it is not exactly
solved in the sense that the generating function of partition
functions has not been explicitly calculated,  in two or three
dimensions. An exactly solved version of the model does exist however
when the restriction of partial directness is imposed on the
configurations of the self-avoiding walk in two dimensions. The model
has been shown to display a tricritical-like collapse transition
\cite{owczarek1993b-:a} as is predicted for the unrestricted model, though with different exponents.

The Interacting Partially Directed Self-avoiding Walk (IPDSAW) model,
and a closely related semi-continuous variant, on the square lattice
was studied extensively in the early 1990's
\cite{binder1990a-:a,brak1992a-a,prellberg1993a-:a,owczarek1993d-:a,owczarek1993b-:a,owczarek1994a-:a}.
It was noticed that this problem is in a family of related problems
including lattice models of vesicles
\cite{brak1994a-:a,prellberg1994a-a} whose solution can be written in
terms of $q$-Bessel functions: moreover, direct correspondences occur
between various models.  Importantly, key work associated with the
asymptotic analysis of the functions that arise in this class of
problems was also completed \cite{prellberg1995d-a}.  Taken together
these works completely solve and analyse the generating function, and
free energy, of the IPDSAW model. In particular, the location of the
collapse transition was found by Binder \emph{et al.\ 
}\cite{binder1990a-:a} while the exact generating function was found
by Brak \emph{et al.\ }\cite{brak1992a-a} in terms of $q$--Bessel
functions. The tricritical nature of the collapse transition was
elucidated by Owczarek \emph{et al.\ }\cite{owczarek1993b-:a} and the
full asymptotics of the generating function can be deduced from the
work of Prellberg \cite{prellberg1995d-a}.

The addition of a stiffness parameter to mimic the effects of
persistence length \cite{grassberger1995a-a} and a stretching
parameter to model the pulling of a polymer by an external force has
more recently been studied in the context of the ISAW model
\cite{grassberger2002a-a}. While a parameter called a pulling force
was not explicitly mentioned in the work on the IPDSAW it was
implicitly part of the set up of the model, as we shall see below,
since the horizontal and vertical steps of the walk were given
separate fugacities. It was shown
\cite{owczarek1993b-:a,prellberg1995d-a} that differentiating the
horizontal and vertical fugacities does not affect the nature of the
collapse transition. Separate analysis of the IPDSAW model with the
force interpretation being explicit confirms this \cite{rosa2003a-a}.
On the other hand, the addition of a stiffness parameter so that the
polymer is now semi-flexible was \emph{not} included in the original
definition of the model. The IPDSAW has recently been reconsidered by
Zhou \emph{et al.\ }\cite{zhou2006a-a}.  Interestingly, they
conjectured, on the basis of Monte Carlo simulation, and an
approximation scheme allowing precise numerical estimates of
thermodynamic quantities, of semi-stiff IPDSAW, that positive
stiffness changes the order of the collapse transition to first-order.
The three-dimensional semi-stiff ISAW model has been studied by
Grassberger and Hegger \cite{grassberger1995a-a} some time ago and
they showed that the collapse transition does indeed become first
order though only for a finite amount of applied stiffness. That is,
small stiffness parameter values do not change the nature of the
collapse transition. A related model shows similar behaviour
\cite{krawczyk2007=a-:a}. In this paper we \emph{solve exactly} the
IPDSAW with stiffness parameter, which we shall now refer to as the
Variably-Flexible Interacting Partially Directed Walk (VFIPDSAW) and
analyse the model in the full parameter space. We show that not only
does the collapse transition become first order when the stiffness
parameter is positive (semi-flexible case) but it is also modified,
though still tricritical, when the stiffness parameter is negative
(super-flexible case).

\subsection{The model}

\begin{figure}[ht!]  
  \centering  
  \includegraphics[width=9cm]{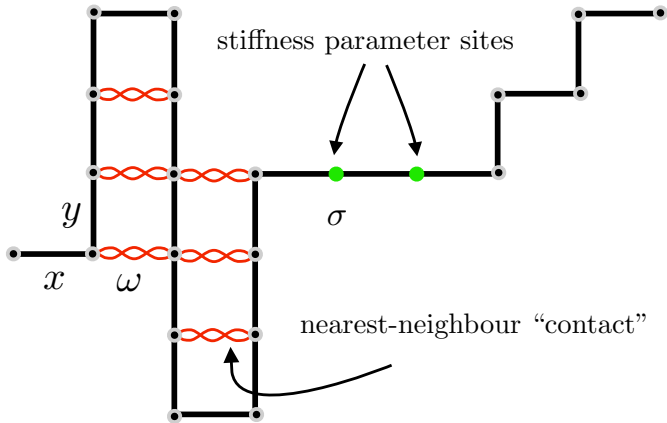}  
  \caption{An example of a Partially Directed Walk of length $L=21$ 
    and horizontal length $N=8$ (the bold black path) with the four
    parameters $x$ associated with horizontal steps, $y$ associated
    with vertical steps, $\omega$ associated with nearest-neighbour
    `contacts' (shown as (red) intertwined curves) and the stiffness
    parameter $\sigma$ associated with the sites (highlighted in grey
    (green)) between two consecutive horizontal steps . The weight of
    the configuration shown is $x^8y^{13}\omega^6\sigma^2$.}
\label{fig-defn-ss-ipdsaw}  
\end{figure}

Consider the square lattice and a self-avoiding walk that has one end
fixed at the origin on that lattice. Now restrict the configurations
considered to self-avoiding walks such that starting at the origin
only steps in the $(1,0)$, $(0,1)$ and $(0,-1)$ are permitted: such a
walk is known as a Partially Directed Self-avoiding Walk (PDSAW). For
convenience, we consider walks that have at least one horizontal step.
Let the total number of steps in the walk be $L$ and the number of
horizontal steps be $N$. Hence, we have $L\geq N\geq 1$.  An example
configuration along with the associated variables of our model are
illustrated in Figure~\ref{fig-defn-ss-ipdsaw}.
We begin by recalling the definition of the IPDSAW model and then add
the ``stiffness parameter'', the addition of which defines our model.

\paragraph{IPDSAW} To define the IPDSAW model we add various energies to properties of this walk and hence
Boltzmann weights to the walk. Firstly, any two occupied sites of the
walk not adjacent in the walk though adjacent on the lattice are
denoted ``nearest-neighbour contacts'': see
Figure~\ref{fig-defn-ss-ipdsaw}.  An energy $-J$ is added for each
such \emph{contact}. We define a Boltzmann weight $\omega=e^{\beta J}$
associated with these contacts, where $\beta = 1/k_BT$ and $T$ is the
absolute temperature. An external horizontal force $f$ pulling at the
other end of the walk adds a Boltzmann weight $p^N$ and $p=e^{\beta f
  a_0}$, with $a_0$ being the length of a lattice bond. The partition
function $Z^{\mathrm{IPDSAW}}_L(\omega,p)$ of the IPDSAW model is
\begin{equation} 
Z^{\mathrm{IPDSAW}}_L(\omega,p) = \sum_{\mathrm{PDSAW\; of\; length\;} L} \omega^m p^N\,,
\end{equation}
where $m$ is the number of nearest-neighbour contacts in the PDSAW. The
generating function $\hat{G}^{\mathrm{IPDSAW}}(z,\omega,p)$ 
\begin{equation}
\hat{G}^{\mathrm{IPDSAW}}(z,\omega,p) = \sum_{L=1}^{\infty} Z^{IPDSAW}_L(\omega,p) z^L \,,
\end{equation}
so $z$ can be considered as fugacity for the steps of the walk and the
generating function a ``generalised partition function'' \cite{owczarek1993b-:a}. 
In previous work \cite{owczarek1993b-:a} an alternate generating function
$G^{IPDSAW}(x,y,\omega)$ was considered, where instead of a force
parameter $p$ horizontal steps were weighted with a fugacity $x$ while
vertical steps were given a fugacity $y$: see
Figure~\ref{fig-defn-ss-ipdsaw}. Hence we have
 \begin{equation}
\hat{G}^{\mathrm{IPDSAW}}(z,\omega,p) = G^{\mathrm{IPDSAW}}(pz,z,\omega)\,.
\end{equation}
Clearly considering a separate horizontal fugacity is equivalent to
considering a horizontal pulling force at the level of generating
functions.

\paragraph{VFIPDSAW} To define the  VFIPDSAW we now add an energy
$-\Delta$ to each site between consecutive horizontal steps of the
walk: see figure~\ref{fig-defn-ss-ipdsaw}. Note that for $\Delta >0$
consecutive horizontal steps are favoured and so this is the positive
stiffness, or semi-flexible, regime while for $\Delta <0$ consecutive
horizontal steps are discouraged so this is the negative stiffness, or
super-flexible regime.  If $\ell$ is the number of such ``stiffness''
sites in a particular IPDSAW then such a configuration is associated
with an additional Boltzmann factor $\sigma^\ell$ where
$\sigma=e^{\beta \Delta}$. That is, each configuration has weight
$\omega^mp^N\sigma^\ell$. Note that one could have equivalently chosen
to weight every bend, or change of direction of the walk, with a
weight $b$ say. That is, each configuration has weight
$\omega^mp^Nb^k$. However, since the number of such bends $k$ is
related to the number of horizontal straight segments, $\ell$,
assuming for convenience at least one vertical step in the walk, as
\begin{equation}
\ell = \frac{(2N - k - 1)}{2}\,,
\end{equation}
then substituting $\sigma=1/b^2$ and setting $p$ to $pb^2$  gives the
same weight for each configuration (barring an overall factor of $b$).

The partition function for VFIPDSAW is defined as
\begin{equation}  
Z^{\mathrm{VFIPDSAW}}_L(\omega,p,\sigma) = \sum_{\mathrm{PDSAW\;of\; length\;} L} \omega^m
p^N \sigma^\ell\,,
\end{equation} 
while the generating function, analogously to the fully-flexible case above, $G^{\mathrm{VFIPDSAW}}(x,y,\omega,\sigma)$, is given as
\begin{equation} 
G^{\mathrm{VFIPDSAW}}(pz,z,\omega,\sigma) = \sum_{L=1}^{\infty} Z^{\mathrm{VFIPDSAW}}_L(\omega,p,\sigma) z^L  \,.
\end{equation}
Clearly one can recover the fully flexible case by setting $\sigma=1$:
\begin{equation}  
G^{\mathrm{IPDSAW}}(pz,z,\omega) = G^{\mathrm{VFIPDSAW}}(pz,z,\omega,1)\,.
\end{equation}

\paragraph{Setup} \emph{As we consider the VFIPDSAW model from now on we shall drop the superscript VFIPDSAW.}
The singularity structure of the generating function as function of $z$ 
determines the free energy.   The reduced free energy is defined as 
\begin{equation} 
\kappa(\omega,p,\sigma) =-\lim_{L\rightarrow\infty} \frac{1}{L} \log
\left[Z_L(\omega,p,\sigma)\right]
\end{equation} 
and is given by 
\begin{equation} 
\kappa(\omega,p,\sigma) = \log z_s (\omega,p,\sigma)\,,
\end{equation} 
where $z_s (\omega,p,\sigma)$ is the closest singularity (on the
positive real axis) of the generating function $G(pz,z,\omega,\sigma)$
in the variable $z$ to the origin.  Note also that
\begin{equation} 
Z_L(\w,p,\sigma)=[z^L]G(pz,z,\w,\sigma)={\frac{1}{2}\pi i}\oint G(pz,z,\w,\sigma){\frac{dz}
  {z^{L+1}}}.  
\end{equation}

In order to find the generating function it is advantageous to rewrite
it in the following way. We can describe the PDSAW configurations in a natural way through the
length $r_i$ of vertical segments between two horizontal steps,
measured in the positive $y$--direction. Each PDSAW begins with a
vertical segment of height $r_1$ followed by an horizontal step. Thus, we associate to each
configuration an $N$--tuple $(r_1,r_2,\ldots,r_N)$ corresponding to a
configuration of total length $L=\sum_{i=1}^N|r_i|+N$.
The energy due to the nearest--neighbour contacts for each of 
these  
configurations is then 
\begin{equation}
-J\;u(r_1,r_2,\ldots,r_N) \,,
\end{equation}
where 
\begin{equation}
u(r_1,r_2,\ldots,r_N)= 
\sum_{i=1}^{N-1}\min(|r_i|,|r_{i+1}|){\Theta}(-r_ir_{i+1})\,, 
\end{equation}
where $\Theta(r)$ is the Heaviside step function:
\begin{equation}
        {\Theta}(r)= \left\{
        \begin{array}{cl}
                0   &  r < 0, \\
                1/2 & r = 0, \\
                1 & r > 0.
        \end{array}     \right. \\
\end{equation}
The number of ``stiffness sites'' $\ell$ is then given by the number
of times $r_i=0$ for any $1<i\leq N$.

We get the generating function by summing  
over all possible lengths as
\begin{equation}
G(x,y,\w,\sigma)=\sum_{L=1}^\infty
\sum_{N=1}^Lx^N\sum_{|r_1|+|r_2|+\ldots+|r_N|=L-N}  
y^{L-N}\w^{u(r_1,r_2,\ldots,r_N)}\sigma^\ell \, ,
\end{equation} 
that is, 
\begin{equation}
 G(x,y,\w,\sigma)= 
\sum_{N=1}^\infty x^N\sum_{M=0}^\infty y^M 
\sum_{|r_1|+|r_2|+\ldots+|r_N|=M}\w^{u(r_1,r_2,\ldots,r_N)}\sigma^\ell\, . 
\end{equation}

\section{Exact solution of the generating function}

In order to derive an expression for $G(x,y,\w,\s)$, 
consider the generalised partition functions $G_r=G_r(x,y,\w,\s)$ 
for walks that start with a vertical segment of height $r$, so that
\begin{equation}
G(x,y,\w,\s)=\sum_{r\in\mathbb{Z}} G_r.
\end{equation}
Then we can concatenate these walks to get a recursion relation 
for $G_r$ as follows:
\begin{equation}
G_r=xy^{|r|}\left\{1+\sigma^{\delta_{r,0}}G_0+\sum_{s\in\mathbb{Z}\setminus\{0\}} \w^{u(r,s)}G_s\right\}\,.
\end{equation}
It follows that
\begin{equation}
G_0=x\left\{1+(\sigma-1)G_0+G \right\}\,,
\end{equation}
so that
\begin{equation}
G_0=xu\left\{1+G \right\}\,,
\label{eqn-g0-bc}
\end{equation}
where 
\begin{equation}
u= \frac{1}{1-x(\sigma-1)}.
\end{equation}
Using the symmetry $G_r=G_{-r}$ and then restricting to $r\geq0$, 
we can further simplify to
\begin{equation}
\label{rec_rel}
G_r=xy^r\left\{1+\sigma^{\delta_{r,0}}G_0+
\sum_{s=1}^\infty (1+\w^{\min(r,s)})G_s\right\}\,,
\end{equation}
which will be the starting point of our investigation. Since $r=0$ is now special we will need to consider $r=1$ separately also:
\begin{equation}
G_1=xy\left\{1+ G_0+(1+\omega) \sum_{s=1}^\infty G_s\right\}\,.
\label{eqn-g1-de}
\end{equation}
Now using 
\begin{equation}
\sum_{s=1}^\infty G_s = \frac{1}{2} (G - G_0)
\end{equation}
gives
\begin{equation}
G_1=xy\left\{1+ \frac{(1-\omega)}{2}G_0+\frac{(1+\omega)}{2} G \right\}\,.
\label{eqn-g1-de2}
\end{equation}
Now using equation~(\ref{eqn-g0-bc}) we obtain
\begin{equation}
G_1=xy\left\{ \frac{(1-\omega)}{2}+\left(\frac{(1+\omega)}{2} + \frac{(1-\omega)}{2} xu\right)(1+G) \right\}\,.
\label{eqn-g1-bc}
\end{equation}

Hence the ratio $G_1/G_0$ can be written in terms of $1+G$. By solving for $(1+G)$ one finds 
\begin{equation}
1+G = \frac{(1-\omega)}{2}\left[ \frac{uG_1}{yG_0} - \left(\frac{(1+\omega)}{2} + \frac{(1-\omega)}{2} xu\right) \right]^{-1}\,.
\label{genfun-g1g0}
\end{equation}

We will now derive a homogeneous second order difference equation 
which we can solve using the same ansatz used previously \cite{owczarek1993b-:a}. Using the 
scaling behaviour of the solutions, we can eliminate one 
of the two linearly independent solutions. We then write the general 
solution of (\ref{rec_rel}) as an expression involving the 
quotient of two $q$--hypergeometric functions.

Taking differences in (\ref{rec_rel}), we first eliminate the 
inhomogeneous term,
\begin{equation}
G_{r+1}-yG_r=\delta_{r,0} xy (1-\sigma) G_0+xq^{r+1}(1-\frac1\w)\sum_{s=r+1}^\infty G_s.
\end{equation}
Here, we introduced for convenience the new variable $q=y\w$.
Upon taking differences a second time, we are left with
\begin{equation}
\label{difference_eqn}
(G_{r+2}-yG_{r+1})-q(G_{r+1}-yG_r)=-\delta_{r,0} qxy (1-\sigma) G_0-xq^{r+2}(1-\frac1\w)G_{r+1}.
\end{equation}
We now solve this equation for $r\geq1$ and subsequently solve for $r=0$.
In the case of no interaction ($\w=1$), the right hand side of 
this equation is zero (for $r\geq1$) and we have a simple
homogeneous difference equation with constant coefficients. 
Its characteristic polynomial $P(\l)$ is
\begin{equation}
P(\l)=(\l-y)(\l-q)
\end{equation}
and the solution is given by $G_r=A_1y^r+A_2q^r$.

This motivates the ansatz \cite{privman1989a-a}
\begin{equation}
\label{ansatz}
G_r=\l^r\sum_{n=0}^\infty q^{nr}c_n\,,
\end{equation}
with $c_n=c_n(x,q,\w)$ independent of $r\geq1$,
which inserted into (\ref{difference_eqn}) gives 
\begin{equation}
P(\l)c_0+\sum_{n=1}^\infty 
q^{nr}\left(P(\l q^n)c_n+xq(1-\frac1\w)\l q^nc_{n-1}\right)=0 .
\end{equation}
This equation is solved by
\begin{equation}
P(\l)=0,\quad\mbox{i.e.}\quad\l_1=y\quad\mbox{and}\quad\l_2=q,
\end{equation}
and, choosing $c_0=1$,
\begin{equation}
c_n=\prod_{m=1}^{n}\frac{-xq(1-\frac1\w)\l q^m}{P(\l q^m)}
=\frac{\left(-x\w(1-\frac1\w)\l\right)^nq^{\binom{n}{2}}}{(\l\w;q)_n(\l;q)_n}.
\end{equation}
Here we have used the standard notation
\begin{equation}
(x;q)_n=\prod_{m=1}^n(1-xq^{m-1}) .
\end{equation}
Defining
\begin{equation}
\label{Hfunction}
H(y,q,t)=\sum_{n=0}^\infty\frac{q^{\binom{n}{2}}(-t)^n}{(y;q)_n(q;q)_n}\,,
\end{equation}
we now can write the general solution of (\ref{difference_eqn}), for $r\geq1$, as
\begin{equation}
G_r=A_1y^rH(y,q,x(1-\frac1\w)q^{1+r})+A_2q^rH(q\w,q,x\w(1-\frac1\w)q^{1+r}).
\end{equation}
We remark that the function $H$ is directly related to a basic
hypergeometric function \cite{gasper1990a-a}
\begin{equation}
H(y,q,t) = {}_{1}\phi_{1} (0,y;q,t) \,,
\end{equation}
which can be seen to be a limiting function of ${}_{2}\phi_1$ and that
is the $q$--deformation of the more familiar hypergeometric function
${}_2F_1$. Analogously, the function $H$ can be understood (apart from
some normalising factors and seen by
taking the limit $q \into 1$) as a $q$--generalisation of Bessel
functions. 
One can easily verify that $H(y,q,t)$ satisfies the following
recurrence
\begin{equation}
qH(y,q,t)+(tq-(y+q))H(y,q,qt) + y H(y,q,q^2t)=0\,.
\label{H-rec}
\end{equation}

Returning to the analysis we see that, for $|q|<1$, $H(y,q,tq^r)$ is
uniformly bounded in $r$, so that we can write
\begin{equation}
|G_r|\leq\const(q^r+y^r).
\end{equation}
This we insert into (\ref{rec_rel}) and, assuming $0<\w^2y<1<\w$ 
we get
\begin{eqnarray}
|G_r|&\leq&\const y^r\left(1+\sum_{s=0}^{r-1}(\w q)^s
+\w^r\sum_{s=r}^\infty q^s\right) \nn \\
&\leq&\const y^r(1+(\w q)^r)\leq\const y^r.
\end{eqnarray}
As $H(y,q,tq^r)\into1$ for $r\into\infty$, we see that in fact
$A_2=0$.
The reason for this is that we obtained the homogeneous difference
equation (\ref{difference_eqn}) by taking differences from
(\ref{rec_rel}),
thus introducing additional solutions.

We now note that the ratios $G_{r+1}/G_r$ contain no unknown constants. In fact, defining
\begin{equation}
{\cal H}(y,q,t)=\frac{H(y,q,qt)}{H(y,q,t)}\,,
\end{equation}
we find
\begin{equation}
\frac{G_{r+1}}{yG_r} = {\cal H}(y,q,x(1-\frac{1}{\omega})q^{1+r})\qquad\text{for $r\geq1$.}
\end{equation}
Note in passing that successive ratios are related via the following recurrence for ${\cal
  H}(y,q,t)$ derived from equation~(\ref{H-rec}),
\begin{equation}
{\cal H}(y,q,t)=q\left[y+q-tq- y {\cal H}(y,q,qt)\right]^{-1}\;.
\label{calH-rec} 
\end{equation} 
In fact, the ratio $G_1/G_0$ is given by a very similar expression. For $r=0$, the recursion (\ref{difference_eqn}) 
can be rewritten as
\begin{equation}
(G_{2}-yG_{1})-q(G_{1}-\frac yuG_0)=-xq^{2}(1-\frac1\w)G_{1}\;,
\end{equation}
from whence one can conclude that
\begin{equation}
\label{G1G0}
\frac{uG_1}{yG_0} = {\cal H}(y,q,x(1-\frac{1}{\omega})q)\;.
\end{equation}

Inserting this ratio into equation~(\ref{genfun-g1g0}) and substituting $q=y\omega$, we have
\begin{equation} 
1+G = \frac{(1-\omega)}{2}\left[ {\cal H}(y,y\omega,x(\omega-1)y) 
 - \left(\frac{(1+\omega)}{2} + \frac{(1-\omega)}{2} xu\right)
\right]^{-1} \,.
\end{equation} 
 We note immediately that the stiffness parameter $\sigma$ only enters (via $u$) in one term of this expression. 
For $\sigma=1$ (i.e. $u=1$), this is exactly equation (4.27) in \cite{owczarek1993b-:a}.

Our final expression for the solution of
the generating function for the \emph{Variably Flexible Interacting
  Partially Directed Walk} in the variables $z,\omega,p,\sigma$ is therefore
\begin{equation} 
1+G = \frac{(1-\omega)}{2}\left[ {\cal H}(z,z\omega,pz^2(\omega-1)) 
 - \left(\frac{(1+\omega)}{2} + \frac{(1-\omega)}{2} \frac{pz}{1-pz(\sigma-1)}\right)
\right]^{-1} \,.
\label{genfun-soln-final} 
\end{equation} 

As we will see below the case $q=1$ is important. From (\ref{calH-rec}) it follows that ${\cal H}(y,1,t)$ is the root
of a quadratic equation, and so
\begin{equation}
{\cal H}(y,1,t)=\frac1{2y}\left[1+y-t-\sqrt{(1+y-t)^2-4y}\right]\;,
\end{equation}
where the branch has been chosen such that ${\cal H}(y,1,t)=1+O(t)$. There is an algebraic singularity at
$(1+t-y)^2=4y$ and the solution is real-valued as long as $(1+t-y)^2\geq4y$.

One can therefore solve for the generating function along the curve $y\omega=1$ as
\begin{equation}  
1+G = \frac{(1-\omega)}{2}\left[{\cal H}(\frac1\omega,1,x(1-\frac1\omega))
 - \left(\frac{(1+\omega)}{2} + \frac{(1-\omega)}{2} xu\right) 
\right]^{-1}  \,.
\label{genfun-soln-q1}  
\end{equation}  
which is now algebraic.

\section{Analysis of the Phase diagram}
\subsection{General considerations}

One can immediately observe that the generating function $G$ has 
singularities at the singularities of 
${\cal H}(y,y\omega,x(\omega-1)y)$ and when the denominator is zero, 
that is, at solutions $\omega_c(x,y,u)$ of
\begin{equation} 
\label{poles}
{\cal H}(y,y\omega,x(\omega-1)y)   
 = \left(\frac{(1+\omega)}{2} + \frac{(1-\omega)}{2} xu\right) \,. 
\end{equation} 
There is an essential singularity of ${\cal
  H}(y,y\omega,x(1-\omega)y)$ when $y\omega=1$. On the other hand,
when the denominator is zero, $G$ has a pole, and the locus
$\omega_c(x,y,u)$ of this pole depends analytically on the parameters
as long as $y\omega<1$. If there is no zero of the denominator for
$y\omega<1$, then the closest singularity is given by the essential
singularity of ${\cal H}(y,y\omega,x(1-\omega)y)$ at $y\omega=1$ where
the generating function converges. On
this curve, we obtain from (\ref{genfun-soln-q1}) that there is an
algebraic singularity at
\begin{equation}
\omega_a(x)=\left(\frac{1+x}{1-x}\right)^2
\end{equation}
and for $u>1$ a simple pole at
\begin{equation}
\omega_p(x,u)=\frac{(1+ux)(1+2x-ux)}{(1-ux)(1-2x+ux)}\;.
\end{equation}
These singularities coincide when $u=1$, at which value the nature of the algebraic singularity changes. 
Note that for $u<1$ the pole disappears.

\begin{figure}[ht!]
\begin{center}
\includegraphics[height=7cm]{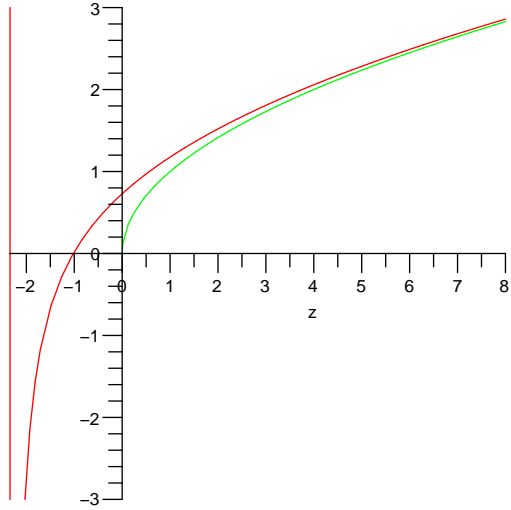}
\caption{\it The graph of $f(z)=-\Ai'(z)/\Ai(z)$. The function $f$ has a simple pole at $z=-2.3381\ldots$, a zero at $z=-1.0187\ldots$,
and is asymptotic to $\sqrt{z}$ for large $z$, which is plotted for comparison.}
\label{airyscaling}
\end{center}
\end{figure}

As stated above for any fixed $\omega$ the generating function as a
function of $y$ either has a pole given by the solution of
$\omega=\omega_c(x,y,u)$ or has a singularity on the curve
$y=1/\omega$. Therefore, for $u\leq1$, $\omega_c(x,y,u)$ meets the algebraic
singularity $\omega_a(x)$ as the curve $y\omega=1$ is approached,
whereas for $u>1$, $\omega_c(x,y,u)$ meets the pole $\omega_p(x,u)$ as
the curve $y\omega=1$ is approached.  To investigate the behaviour of
$G$ near the curve $y\omega=1$ more closely, we need the asymptotic
behaviour of ${\cal H}(y,q,t)$ as $q\rightarrow1$. Away from the
algebraic singularity, i.e. for $(1+t-y)^2>4y$, we can use
(\ref{calH-rec}) to derive an asymptotic expansion in $\epsilon=1-q$,
\begin{equation}
{\cal H}(y,1-\epsilon,t)\sim\sum_{n=0}^\infty{\cal H}^{(n)}(y,t)\epsilon^n
\end{equation}
with the first terms given by ${\cal H}^{(0)}(y,t)={\cal H}(y,1,t)$ and
\begin{align}
\label{Hexpansion}
{\cal H}^{(1)}(y,t)=&\frac{{\cal H}^{(0)}(y,t)}{y{\cal H}^{(0)}(y,t)^2-1}
\left[1+yt{\cal H}^{(0)}(y,t)\frac\partial{\partial t}{\cal H}^{(0)}(y,t)+(t-1){\cal H}^{(0)}(y,t)\right]\\
=&
-\frac1{2y}\left(1+\frac{y+t-1}{\sqrt{(1+y-t)^2-4y}}+\frac{2yt}{(1+y-t)^2-4y}\right)
\;.
\end{align}
Close to the algebraic singularity at $q=1$, the singularity structure is significantly more complicated, but has been thoroughly
elucidated in \cite{prellberg1995d-a}. Using Lemma 4.3 from \cite{prellberg1995d-a}, a result completely analogous to
Theorem 5.3 in \cite{prellberg1995d-a} can be obtained for ${\cal H}(x,q,t)$, i.e. an asymptotic expansion in $q=1-\epsilon$ 
uniformly valid for all values of $t$ and $x$, which reads
\begin{equation}
{\cal H}(y,1-\epsilon,t)=\frac1{2y}\left[1+y-t-
\left(-\frac{\Ai'(\alpha\epsilon^{-2/3})}{\alpha^{1/2}\epsilon^{-1/3}\Ai(\alpha\epsilon^{-2/3})}\right)
\sqrt{(1+y-t)^2-4y}\right](1+O(\epsilon))\;.
\label{Hasy}
\end{equation}
Note that for $\epsilon\rightarrow0$ the expression multiplying the square root in (\ref{Hasy}) tends to $1$ as is necessary.
Here, $\alpha=\alpha(y,t)$ is a function of $y$ and $t$ which is known exactly \cite{prellberg1995d-a}. While the precise 
form of $\alpha$ is rather cumbersome, it simplifies considerably near the critical point, and we find 
\begin{equation}
\label{alpha}
\alpha(y,t)\sim\left(\frac4{1-(t-y)^2}\right)^{4/3}\frac{(1+y-t)^2-4y}4
\end{equation}
for small $(1+y-t)^2-4y$. This implies that here
\begin{equation}
{\cal H}(y,1-\epsilon,t)\sim\frac1{2y}\left[1+y-t+\epsilon^{1/3}
\frac{\Ai'(\alpha\epsilon^{-2/3})}{\Ai(\alpha\epsilon^{-2/3})}
\frac{(1-(t-y)^2)^{2/3}}{2^{1/3}}
\right]\;.
\label{Hasy2}
\end{equation}
The behaviour of this expression is determined by the function $f(z)=-\Ai'(z)/\Ai(z)$, the graph of which is shown in figure 
\ref{airyscaling}. The large-$z$ asymptotics allows for matching for $\epsilon\rightarrow0$ and positive $\alpha$.
For negative $\alpha$, the argument of $f$ is negative. As $f(z)$ has a simple pole at $z=-2.3381\ldots$, for any fixed $\alpha<0$
we have a pole at a finite value of $\epsilon$. As $\alpha$ tends to zero, the locus of this pole scales as $\epsilon^{2/3}$.

\subsection{Fully flexible case ($\Delta=0$)} 

This case is the one considered in our earlier work \cite{owczarek1993b-:a}.
Here $u=1$, and the simple pole at $\omega_c(x,y,1)$ approaches an algebraic singularity at 
$\omega_t=\left(\frac{1+x}{1-x}\right)^2>1$, at which the generating function \textit{diverges}.
In particular, we find
\begin{equation}
\label{special}
1+G\sim\left(-\frac{\Ai'(\alpha\epsilon^{-2/3})}{\alpha^{1/2}\epsilon^{-1/3}\Ai(\alpha\epsilon^{-2/3})}\right)^{-1}
\frac{\omega-1}{\sqrt{(1+\omega+x(1-\omega))^2-4\omega}}
\end{equation}
as $\epsilon=1-y\omega\rightarrow0$, with $\alpha=\alpha(\frac1\omega,x(1-\frac1\omega))$ given by (\ref{alpha}). 
Near the transition, $\alpha$ is small and we can write
\begin{align}
1+G\sim&-A\epsilon^{-1/3}\frac{\Ai(\alpha\epsilon^{-2/3})}{\Ai'(\alpha\epsilon^{-2/3})}
\end{align}
with $A=(\omega-1)/[\omega^2-(x(\omega-1)-1)^2]^{2/3}(2\omega)^{1/3}$.
For $\omega=\omega_t$, $\alpha=0$ and $G$ diverges as $\epsilon^{-1/3}$. For $\omega>\omega_t$, $\alpha>0$ and $G$ tends to a
finite value given by (\ref{genfun-soln-q1}). For $\omega<\omega_t$, $\alpha<0$ and $G$ has a simple pole 
at $\alpha\epsilon^{2/3}\approx-1.0187\ldots$.
We accordingly have a second-order phase transition characterised by
\begin{equation}
\gamma_u=\frac12\qquad\gamma_t=\frac13\qquad\phi=\frac23\;,
\end{equation}
with the exponents as defined by Owczarek \emph{et al.\ }\cite{owczarek1993b-:a}.

Changing to the variables $z,\omega,p,\sigma$ this can be formulated as follows. This is the case $\sigma=1$, and there is a curve 
of simple poles given by
$\omega_c(pz,z,1)$ approaching the curve $\omega z=1$ at $\omega_t$ given by $\omega_t=\left(\frac{p+\omega_t}{p-\omega_t}\right)^2$. 
For $p=1$, the solution is $\omega_t=3.3829\ldots$. 
Near the transition, (\ref{special}) holds with the appropriate substitution
$x=pz$ and $y=z$. While the location $\omega_c(pz,z,1)$ of the poles, as well as $\omega_t$ change as a function of $p$, the 
character of the phase transition does not.

\subsection{Super-flexible case ($\Delta < 0$)}

Now $u<1$, and the simple pole at $\omega_c(x,y,u)$ approaches an algebraic singularity at 
$\omega_t=\left(\frac{1+x}{1-x}\right)^2>1$, at which the generating function \textit{converges}.
In particular, we find
\begin{align}
\label{less-special}
1+G\sim&\frac{\omega-1}{x(\omega-1)(1-u)+\left(-\frac{\Ai'(\alpha\epsilon^{-2/3})}{\alpha^{1/2}\epsilon^{-1/3}\Ai(\alpha\epsilon^{-2/3})}\right)
\sqrt{(1+\omega+x(1-\omega))^2-4\omega}}
\end{align}
as $\epsilon=1-y\omega\rightarrow0$, with $\alpha=\alpha(\frac1\omega,x(1-\frac1\omega))$ given by (\ref{alpha}). 
Near the transition, $\alpha$ is small and we can write
\begin{align}
1+G\sim&\frac1{x(1-u)-A^{-1}\epsilon^{1/3}\frac{\Ai'(\alpha\epsilon^{-2/3})}{\Ai(\alpha\epsilon^{-2/3})}}
\end{align}
with $A$ as above.
For $\omega=\omega_t$, $\alpha=0$ and $G$ converges with the singular part scaling as $\epsilon^{1/3}$. 
For $\omega>\omega_t$, $\alpha>0$ and $G$ tends to a
finite value given by (\ref{genfun-soln-q1}). For $\omega<\omega_t$, $\alpha<0$ and $G$ has a simple pole 
at some value of $\epsilon$ where $-\Ai'(z)/\Ai(z)<0$, i.e. $-2.3381\ldots<\alpha\epsilon^{2/3}<-1.0187\ldots$.
We now find a second-order phase transition with
\begin{equation}
\gamma_u=-\frac12\qquad\gamma_t=-\frac13\qquad\phi=\frac23\;.
\end{equation}
As $u\rightarrow1$, we recover the fully flexible case discussed above.

Changing to the variables $z,\omega,p,\sigma$ this can be formulated as follows. This is the case $\sigma<1$, and there is a curve 
of simple poles given by
$\omega_c(pz,z,(1-pz(\sigma-1))^{-1})$ approaching the curve $\omega z=1$ at $\omega_t$ given by $\omega_t=\left(\frac{p+\omega_t}{p-\omega_t}\right)^2$. 
This transition point therefore is independent of the value of $\sigma$, and is identical to the one obtained in the fully flexible case.
Near the transition, (\ref{less-special}) holds with the appropriate substitution
$x=pz$, $y=z$, and $u=1/(1-pz(\sigma-1))$. While the location $\omega_c(pz,z,(1-pz(\sigma-1))^{-1})$ of the poles, as well as 
$\omega_t$ change as a function of $p$, the character of the phase transition does not.

\subsection{Semi-flexible case ($\Delta > 0$)}

Now $u>1$, and the simple pole at $\omega_c(x,y,u)$ approaches a simple pole at 
$\omega_t=\frac{(1+ux)(1+2x-ux)}{(1-ux)(1-2x+ux)}>1$.
In particular, we find near the transition that
\begin{equation}
\label{not-special}
1+G\sim\frac{\omega-1}{x(\omega-1)(1-u)+\sqrt{(1+\omega+x(1-\omega))^2-4\omega}-2\epsilon{\cal H}^{(1)}(\frac1\omega,x(\omega-1)\frac1\omega)}
\end{equation}
as $\epsilon=1-y\omega\rightarrow0$, with ${\cal H}^{(1)}$ given by (\ref{Hexpansion}). Note that 
$x(\omega-1)(1-u)+\sqrt{(1+\omega+x(1-\omega))^2-4\omega}$ is asymptotically linear in $\omega-\omega_t$ and is negative for
$\omega<\omega_t$. Note further that ${\cal H}^{(1)}<0$.
For $\omega>\omega_t$, $G$ tends to a finite value as $\epsilon\rightarrow0$. This expression diverges with a simple pole
as $\omega$ approaches $\omega_t$. Similarly, for $\omega=\omega_t$, $G$ diverges as $\epsilon^{-1}$.
For $\omega<\omega_t$, $G$ has a simple pole at 
\begin{equation}
\epsilon=\frac{x(\omega-1)(1-u)+\sqrt{(1+\omega+x(1-\omega))^2-4\omega}}{2{\cal H}^{(1)}(\frac1\omega,x(\omega-1)\frac1\omega)}\;.
\end{equation}
We accordingly find a first-order phase transition with
\begin{equation}
\gamma_u=1\qquad\gamma_t=1\qquad\phi=1\;.
\end{equation}
As $u\rightarrow1$, ${\cal H}^{(1)}$ diverges, and the character of the phase transition changes as characterised above.

Changing to the variables $z,\omega,p,\sigma$ this can be formulated as follows. This is the case $\sigma>1$, and there is a curve 
of simple poles given by
$\omega_c(pz,z,(1-pz(\sigma-1))^{-1})$ approaching the curve $\omega z=1$ at $\omega_t$ given by 
\begin{equation}
\omega_t=\frac{(2(1-\sigma)p^2+(2-\sigma)\omega_tp+\omega_t^2)((2-\sigma)p+\omega_t)}
{(2(\sigma-1)p^2-\omega_tp\sigma+\omega_t^2)(\omega_t-p\sigma)}\;.
\end{equation}
While this is quite cumbersome in general, some special values have simple solutions. For example, at $p=1$ and $\sigma=2$ we find
$\omega_t=2+\sqrt2=3.4142\ldots$.
Near the transition, (\ref{not-special}) holds with the appropriate substitution
$x=pz$, $y=z$, and $u=1/(1-pz(\sigma-1))$. While the location $\omega_c(pz,z,(1-pz(\sigma-1))^{-1})$ of the poles, as well as 
$\omega_t$ change as a function of $p$, the character of the phase transition does not.

\section{Conclusion}

We have analysed the exact solution of a two-dimensional lattice model
of a single polymer in solution containing parameters that vary the
intra-polymer attraction, the amount of horizontal stretching force
applied and the amount of stiffness. The restriction of partial
directness is required to ensure solvability. We find that a
tricritical collapse transition takes place for no stiffness or
negative stiffness, and that this is unaffected by an horizontal
force. The entropic exponents are different in the negative stiffness
regime to those in the zero stiffness regime. On the other hand, when
the polymer becomes semi-stiff the collapse transition immediately
becomes first-order. 

\section*{Acknowledgements}
Financial support from the Australian Research Council via its support
for the Centre of Excellence for Mathematics and Statistics of Complex
Systems is gratefully acknowledged by the authors.


\end{document}